\documentclass[twocolumn,showpacs,preprintnumbers,amsmath,amssymb,prc]{revtex4}

\usepackage{graphicx}
\usepackage{dcolumn}
\usepackage{bm}

\usepackage{color}
\usepackage{CJK}
\usepackage[colorlinks,linkcolor=blue,urlcolor=blue,citecolor=blue]{hyperref}

\begin{document}
\begin{CJK*} {UTF8} {gbsn}

\title{Dipole excitation of $^6$Li and $^9$Be studied with an extended quantum molecular dynamics model}

\author{Bo-Song Huang(黄勃松)\footnote{Email: huangbosong@sinap.ac.cn}}
\affiliation{Shanghai Institute of Applied Physics, Chinese Academy of Sciences, Shanghai 201800, China}

\author{Yu-Gang Ma(马余刚)\footnote{Email: mayugang@fudan.edu.cn}}
\affiliation{Key Laboratory of Nuclear Physics and Ion-Beam Application (MOE), Institute of Modern Physics, Fudan University, Shanghai 200433, China}

\begin{abstract}
The $\alpha$ ($^4$He) - clustering structure is a common phenomenon in light nuclei due to the decreasing contribution of mean field in few-body system. In this work we presented calculations of giant dipole resonance (GDR) excitations for two non $\alpha$-conjugate light nuclei, namely $^{6}$Li and $^{9}$Be, within a framework of an extended quantum molecular dynamics model. For $^{6}$Li, we investigated the GDR spectra from
the two-body clustering structure with  $\alpha$ + deuteron  as well as the  three-body structure with $\alpha$ + $n$ + $p$, and  found that the major  $\alpha$-clustering contribution on the GDR peak is located at around  31 MeV,
while the resonance contributions between clusters, namely $\alpha$ and deuteron or ($n$ + $p$), are located on the lower energy side,  which can be regarded as  pygmy dipole resonance (PDR). For  $^{9}$Be,
a mixture  configuration contribution for  the Chain-like structure and Borromean-like structure of the $\alpha$ + n + $\alpha$ configuration can somehow explain its GDR results.
\end{abstract}


\maketitle

\section{Introduction}

The phenomena of $\alpha$-clustering structure in light nuclei have been investigated by lots of experiments and theories for a few decades, it is not only of  importance to understanding effective nucleon-nucleon interaction  inside nuclei, but also help to learn their important roles during the nucleosynthesis processes
\cite{burbidolde1957,kappeler2011,Hoyle,Ikeda,Ortzen,THSR,Freer,Nature,ZhouB,YangZH,An2,Tang,LiWJ,Liu1,Zhang,Liu2}.

For  light nuclei whose binding  energy is weak,
the mean field effect is not strong enough to break cluster structure, therefore the clustering behavior could be observed in the excited states or even in the ground state. 
 Some probes were presented to explore sensitivity to the clustering structure. For instance, the collective observables show significant
difference among various $\alpha$-clustering structures in heavy-ion collisions \cite{Guo,ZhangS,ZhangS2,XuZW,YeYL}, and nucleon-nucleon correlation displays different behavior  in  three-body photodisintegration  of $\alpha$-clustering nuclei \cite{huang12C,huang16O,HuangBS-2019,Zhou}.  In addition, another important  probe, so-called giant dipole resonance (GDR) spectrum shows its sensitivity  to  different  configurations of $^{12}$C and $^{16}$O  in a framework of an extended quantum molecular dynamics (EQMD)  model \cite{W.B.He,W.B.He2}. However, the above GDR results were only investigated from the $\alpha$-conjugate light nuclei, and are not yet checked for non $\alpha$-conjugate light nuclei within the same model. Based upon this motivation, we performed such investigation for  $^6$Li and $^9$Be nuclei in this work. On the other hand,  it is noted that some studies for dipole excitation of the non $\alpha$-conjugate light nuclei, such as $^9$Be \cite{Arai1996,Arai2003,Kikuchi1,Kikuchi2} and $^6$Li \cite{Arai1995} as well as its  mirror nucleus $^6$He \cite{Arai1999}  have been performed with  some microscopic cluster models by using three-cluster model or the extended three-cluster model. In addition, the  isovector giant dipole states  in the continuum for $^{16}$O and even-even Be isotopes were studied by the continuum random-phase approximation \cite{Nakatsukasa1}.  

Giant dipole resonance,  as one of the most noticeable nuclear collective motions, has been extensively discussed in low energy nuclear reactions. It was recognized as the classical picture of the oscillation between protons and neutrons induced by the $E1$ external field. Many studies have been performed experimentally and theoretically for this kind of collective excitation, and some nice review papers can be found in literatures, eg. Refs.~\cite{Rev1,Rev2,Rev3}.

In recent experiments, the study of GDR for $^{6}$Li was performed   by measuring the total absorption cross-section of  $^{6}$Li$(\gamma,$xn$)$ reactions,
and it shows the two-component contribution with a low and a high energy peak at around 12 MeV and 33 MeV, respectively~\cite{T. Yamagata}. Theoretically, the $E1$ transition of $^{6}$Li was studied with a fully microscopic six-body calculation,  in which the final state was described by three types of configuration, i.e., single particle excitation, $\alpha$ + $p$ + $n$ as well as $^3$He + $t$. The conclusion is that  the $E1$ excitation is dominated by the $^3$He  + $t$ configuration at low energy around 20 MeV, while it has a mixed contribution and competition between the $\alpha$ + $p$ + $n$
 and  $^3$He  + $t$  configuration above 30 MeV \cite{S. Satsuka}.

 As for $^{9}$Be, experimentally,  significant strength of the cluster dipole resonance at about 10 MeV has been measured in comparison with the GDR through quasi-monochromatic $\gamma$-ray by photodisintegration \cite{H.Utsunomiya}. However,  one can not distinguish that the structure of $^{9}$Be is whether $^{8}$He + $n$ or $\alpha$ + $n$ + $\alpha$. Theoretically, the low-lying states of $^9$Be are understood by 2$\alpha$ + $n$ cluster structure as discussed in cluster
models  \cite{Oka1,Oka2,Fon,Des}. The Antisymmetrizied Molecular Dynamics (AMD) model was used to calculate the low energy isovector dipole excitations. The ground state is used by the so-called sAMD + $\alpha$GCM, where sAMD is the ``shifted basis AMD" by shifting the position of the Gaussian centroid of the $i$-th single-particle wave function, and GCM represents the generator coordinate method.  The angular-momentum and parity projection are taken into account for excited states  for $^{9}$Be and the structure of two-$\alpha$ clusters core plus a valence neutron is obtained, and a conclusion that the first lower energy peak is mainly contributed by the longitudinal mode and the higher energy peak comes from the transverse mode was drawn in Ref.~\cite{Yoshiko}.

Although $^{6}$Li has been generally explained as the predominant $\alpha$ + deuteron clustering structure \cite{S. Satsuka}, however, it is not clear that how its  possible configuration of nuclear clustering structure contributes on the excitation of GDR.  For $^{9}$Be,  whether  it is a configuration of two $\alpha$s and a neutron is also not clear.
In order to understand the above exotic light nuclei,  namely $^{6}$Li and $^{9}$Be,  we calculate their GDR spectra by assuming  different initial structures in this work in a framework of EQMD and compare the calculated GDR spectra with the experimental data.

The rest of  paper is organized as follow:   Section II provides a brief introduction of the EQMD model and describe the methods of calculation of GDR spectra. In section III we discuss the calculation results together  with the experimental data for the GDR spectra. Finally, a summary is given in Section IV.

\section{Model and methodology}

\subsection{EQMD model introduction }

Quantum molecular dynamics (QMD) type models~\cite{J.Aichelin,J.Aichelin2} have been extensively applied for describing reaction dynamics and  fragment formation  in heavy ion collisions at intermediate energy \cite{C.Hartnack,C.Hartnack1,FengNST,Yan-NST1,Yan-NST2}.
EQMD model ~\cite{MARUYAMA} is one of  extension versions of  QMD model, in which the description of the ground state of the nuclear system has been significantly improved by obtaining  the lowest point of energy of the nuclei through the cooling process which cancels the zero-point energy caused by the wave packet broadening in the standard QMD. On the other hand, repulsion between identical nucleons is
phenomenologically taken into account by a repulsive Pauli potential ~\cite{A.Ohnishi}.
As a result, saturation property and cluster structure can be obtained after energy cooling in the EQMD
model  \cite{W.B.He,W.B.He2}. Different from the traditional QMD model ~\cite{J.Aichelin,J.Aichelin2},  the width of each wave packet in the  EQMD model is taken as a dynamical variable~\cite{P.Valta}  and the wave packet of the nucleon with the form of Gaussian-like as 
\begin{multline}
\phi_{i}(r_{i}) = \bigg(\frac{v_i+v^{*}_{i}}{2\pi}\bigg)^{3/4}exp\bigg[-\frac{v_{i}}{2}(\vec{r}_{i}-\vec{R}_{i})^{2} +\frac{i}{\hbar}\vec{P}_{i}\cdot \vec{r}_{i}\bigg],
\end{multline}
where $\vec{R}_{i}$ and $\vec{P}_{i}$ are the centers of position and momentum of the $i$-th wave packet, and the $v_{i}$ is the width of wave packets which can be presented as
${v_i} = {{1/{\lambda _i}}} + i{\delta _i}$  where $\lambda_i$ and $\delta_i$ are dynamical variables.
The ${v_i}$ of Gaussian wave packet for each nucleon is dynamical and independent.

The Hamiltonian of the whole system is written as Eq. (2)
\begin{multline}
H = \left\langle \Psi \mid \sum_{i} -\frac{\hbar^{2}}{2m}\bigtriangledown^{2}_{i}-\widehat{T}_{c.m.}+\widehat{H}_{int} \mid\Psi \right\rangle\\
\\=\sum_{i}\bigg[\frac{{\vec{P}_i}^2}{2m}+\frac{3\hbar^{2}(1+\lambda^{2}_{i}\delta^{2}_{i} )}{4m\lambda_{i}} \bigg]-T_{c.m.}+H_{int},
\label{eq_H}
\end{multline}
where $T_{c.m.}$ is the zero-point center-of-mass kinetic energy due to the zero-point oscillation of center-of-mass motion, which appears because center-of-mass wave function is fixed to be a Gaussian wave packet in the model  \cite{MARUYAMA}.  $\Psi$ is the direct product of Gaussian wave packets of
nucleons and
$H_{int}$  is the interaction  potential with the form of
\begin{equation}
H_{int} = H_{Skyrme} + H_{Coulomb} + H_{Symmetry} + H_{Pauli},
\end{equation}
where the Pauli potential $H_{Pauli} = \frac{c_{ P}}{2}\sum_{j}(f_{i}-f_{0})^{\mu}\theta(f_{i}-f_{0})$
with $f_{i}$ is defined as an  overlap of $i$-th nucleon with other nucleons which have the same spin and isospin.

The zero-point center-of-mass kinetic energy shown in right side of Eq.~(2) does not cause  trouble for some nuclear information, however, may cause a serious trouble in treating fragment formation, and should be subtracted. In this case,   the zero-point center-of-mass kinetic energy of the system $T_{c.m.}$ is subtract following the basic idea of Ref.~\cite{Ono}. In EQMD case, all the wave packets have different contributions to zero-point kinetic energy. $T_{c.m.} = \sum_i\frac{t^{c.m.}_i}{M_i}$, where ${t^{c.m.}_i}$ is the zero-point kinetic energy of the wave packet $i$ written as 
\begin{equation}
t^{c.m.}_i = -\frac{<\phi_i |  \hbar^2 \bigtriangledown^2 | \phi_i >}{2m} + \frac{<\phi_i |\hbar \bigtriangledown | \phi_i>^2}{2m},
\end{equation}
and $M_i$ is the ``mass number" of the fragment to which the wave packet $i$ belongs. The ``mass number" is calculated as the sum of the ``friendships" the nucleons
\begin{equation}
     M_i = \sum_j {F_{ij}},
\end{equation}
 where
\begin{equation}
F_{ij} = \begin{cases}
1 & {(|\bold{R_i}-\bold{R_j}|<a)}
\\e^{-(|\bold{R_i}-\bold{R_j}|-a)^2/b} & {(|\bold{R_i}-\bold{R_j}|\geq a)},
\end{cases}
\end{equation}
where $a$ and $b$ are parameters for describing cluster formation. Usually, $a$ = 1.7 fm and $b$ = 4.0 fm$^2$ are taken in treating $\alpha$-conjugate clustering nuclei \cite{MARUYAMA}, but they need to be adjusted in present work for non $\alpha$-conjugated nuclei, which will be discussed in Sec. III A. 

Using the above model, we can easily obtain  $\alpha$-clustering structures for $\alpha$-conjugate nuclei, such as $^8$Be, $^{12}$C and $^{16}$O in our previous work \cite{W.B.He,W.B.He2}.
However, in this work, we focus on the studies of non $\alpha$-conjugated clustering light nuclei which are used for further analysis of their GDR spectra.

\subsection{GDR algorithm}

The giant dipole resonance can be considered as the classical picture of the oscillation between the bulk of protons and neutrons
along the opposite direction inside the excited nucleus based on the Goldhaber-Teller assumption ~\cite{Goldhaber},  and therefore the oscillation energy spectra can be calculated. Other method, such as  the random-phase approximation (RPA),  solves  through linear response by density-functional theory  as shown in Ref.~\cite{Nakatsukasa,Nakatsukasa1,Cao}.
The isovector giant dipole moment in coordinator space $D_{R}(t)$ and in momentum space $D_{K}(t)$ can be written as follows, respectively~\cite{Baran, H. L. Wu,C. Tao,Tao2,Ye,GuoCQ,Wang1,Wang2}
\begin{equation}
 D_{G}(t) = \frac{NZ}{A}[R_{Z}(t)-R_{N}(t)],
\end{equation}
and
\begin{equation}
 K_{G}(t) = \frac{NZ}{Ah}[\frac{P_{Z}(t)}{Z}-\frac{P_{N}(t)}{N}],
\end{equation}
where $R_{Z}(t)$ and $R_{N}(t)$ are the center of mass of the protons and neutrons in coordinate space, respectively,  and $P_{Z}(t)$ and $P_{N}(t)$ are the center of mass of the protons and neutrons in momentum space, respectively.

Through the dipole moment $D_{G}(t)$, the strength of  dipole resonance or the $\gamma$ emission probability of the system at energy $E_{\gamma} = \hbar \omega$ can be derived from the following formula:
\begin{equation}
 \frac{dP}{dE_\gamma} = \frac{2e^{2}}{3\pi \hbar c^{3}E_\gamma}|D^{''}(\omega)|^{2},
 \label{Eg}
\end{equation}
where $\frac{dP}{dE_{\gamma}}$ can be  interpreted as the $\gamma$ emission probability,  and $D^{''}(\omega)$  means the Fourier transformation of the second derivative of $D_{G}(t)$ with respect to time:
\begin{equation}
D^{''}(\omega)^{2} = \int_{t_{0}}^{t_{max}} D_{G}^{''}(t)e^{i\omega t}dt.
\end{equation}
For oscillation between clusters such as $\alpha$ + deuteron for $^6$Li, the calculation for its strength is just similar to above equations, therein $D_{G}(t)$ is proportional to the distance between the two centroids of clusters, then the spectral function with  $E_{\gamma} = \hbar \omega$ can be obtained by using Eq.~(\ref{Eg}).

\section{Results and discussion}

\subsection{Initial configuration of $^6$Li and $^9$Be}

\begin{figure}
\center
\includegraphics[scale=0.4]{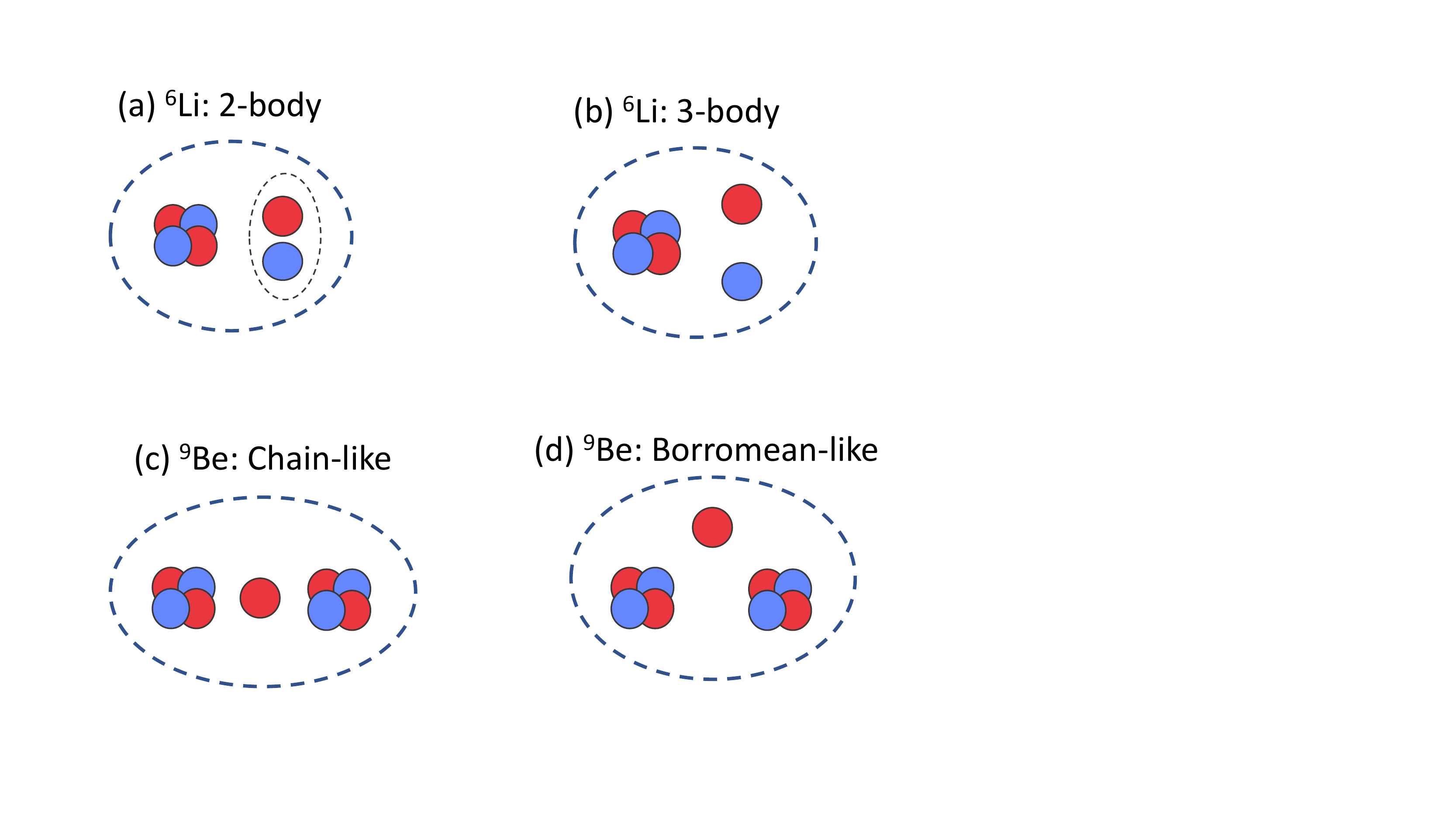}
\vspace{-1.0cm}
\caption{The initial configurations of $^{6}$Li and $^{9}$Be from the EQMD calculations. The upper panels depict two types of clustering structures of $^{6}$Li, i.e. the 2-body structure of $\alpha$ + deuteron (a) as well as the 3-body structure of $\alpha$ + $n$ + $p$ (b). The lower panels show 3-body clustering structures of $^{9}$Be,  i.e. one is the Chain-like configuration with a neutron in between two $\alpha$s (c) and another is the Borromean-like configuration with two-$\alpha$s +  a neutron (d).  }
\label{fig_6Li-shapes}
\end{figure}

Before we calculate the strength of the GDR spectra, we need to assume the initial configurations of those light nuclei. To this end,   two different $\alpha$-clustering configurations of $^6$Li and $^9$Be are obtained through the cooling process within the EQMD model.
 As discussed  in Sec.IIA, there are two parameters of $a$ and $b$ in the present model related to the  zero-point kinetic energy.
A configuration which we required for $^{6}$Li was achieved by hand with an $\alpha$ and a pair of neutron and proton.   In order to keep stability of the configuration for these nuclei, we take $a$ = 1.0 fm and $b$ = 1.0 fm$^2$  for  a neutron-proton pair, 
$a$ = 1.5 fm and $b$ = 1.41 fm$^2$ for $\alpha$ cluster and  $a$ = 1.5 fm and $b$ = 1.0 fm$^2$ for the nucleon between the 
 a neutron-proton pair and $\alpha$. In contrast,  $a$ = 1.7 fm and $b$ = 4.0 fm$^2$ was usually used in previous EQMD calculations. Here, the zero-point kinetic energy plays the role of repulsion, and its subtraction reduces the repulsive interaction between the two clusters.  After the cooling process with the above $a$ and $b$ parameters, two types of clustering configurations emerge. The upper panels of Fig.~\ref{fig_6Li-shapes} show that $^6$Li is either presented as the 2-body clustering structure of  $\alpha$ + deuteron  (Fig.~\ref{fig_6Li-shapes}(a)) or as a 3-body  clustering structure of $\alpha$ + $p$ + $n$ (Fig.~\ref{fig_6Li-shapes}(b)).  Table I shows the binding energies per nucleon of $^6$Li ($E_{b}$), $\alpha$-cluster (${E_\alpha}$),   $d$ or $n + p$ ($E_{d(np)}$)  calculated from the EQMD model along with a comparison to experimental data. ${E_\alpha}$, $E_{d(np)}$ and $E_{np}$ were obtained  by the sum of nucleon binding energy inside the $\alpha$-cluster, inside deuteron  or  $n+p$, respectively.  For the $\alpha$ + $d$ configuration, its experimental data of excitation energy is 1.47 MeV \cite{T. Yamagata} , while for $\alpha$ + $n$ + $p$ it is 3.7 MeV \cite{T. Yamagata}, thus the binding energy of the two configurations is -5.087 MeV and -4.715 MeV, respectively, considering the ground state binding energy of $^6$Li. By comparing with the data, our model calculation results  seem slightly smaller, which can be attributed that the EQMD model can not describe energy level structure accurately. Nevertheless, the trend is generally consistent with the data.   For $^9$Be,  the three-body clustering structures of $\alpha$ + $n$ + $\alpha$ appear: the one is the Chain-like structure as shown in Fig.~\ref{fig_6Li-shapes}(c) and the another is  the Borromean-like  configuration as shown in  Fig.~\ref{fig_6Li-shapes}(d).

\begin{figure}
\center
\includegraphics[scale=0.45]{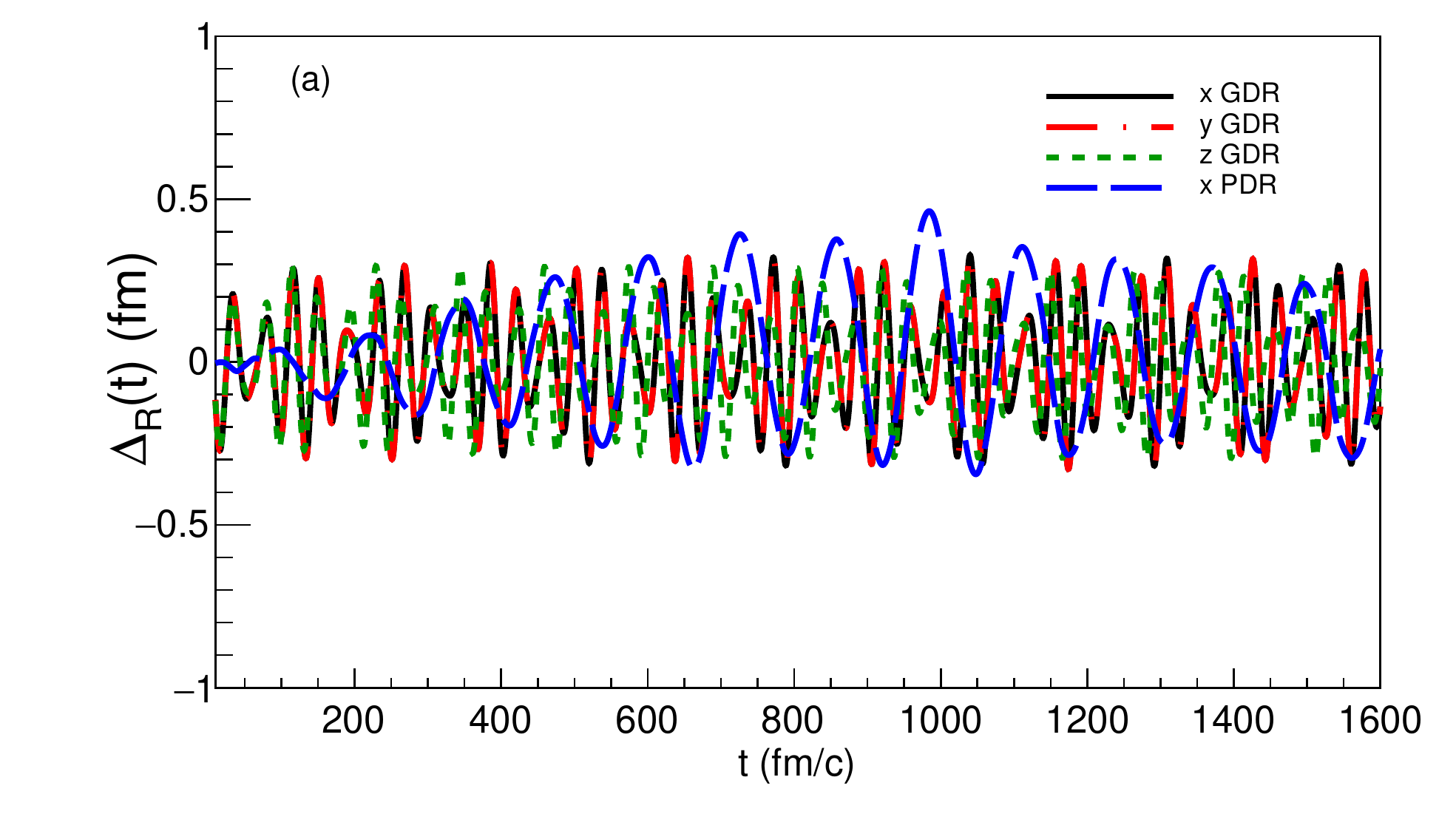}
\includegraphics[scale=0.45]{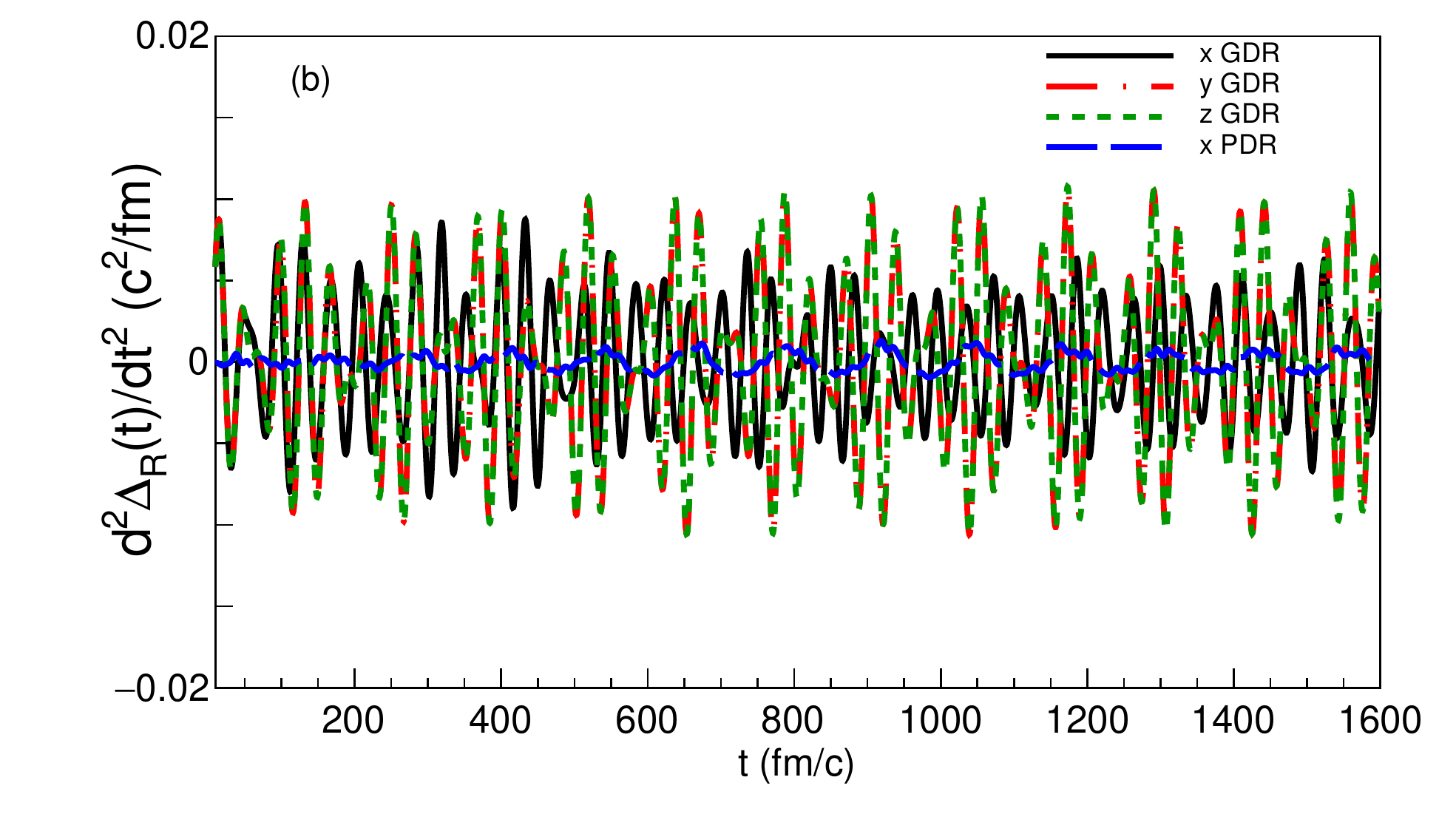}
\caption{ The time evolution of the moments $\Delta_R(t)$ = $R_Z(t) - R_N(t)$ or $R_{\alpha}(t) - R_{d}(t)$ 
in coordinate space (a) and  their second derivatives
with respect to time (b) for the $\alpha$ + $d$ configuration of $^6$Li. 
In the insert, $x$-, $y$-
 and $z$-GDR represent three components of excited dipole moments along $x$, $y$ and $z$ directions of the oscillation of proton's core against  neutron's core, and while  the blue long-dashed line (marked as ``x PDR" in the insert) represents the oscillation between the core of $\alpha$ against the core of deuteron.}
\label{fig_time}
\end{figure}

\begin{table}[]
\centering
\caption{Binding energies  per nucleon of $^6$Li ($E_{b}$), $\alpha$-cluster (${E_\alpha}$), $d$ or $n + p$ ($E_{d(np)}$) calculated  for two configurations from the $^6$Li cooling process of  EQMD model.
The data $E_{exp}$  was calculated by the total ground-state binding energy minus excitation energy which was taken from the experimental data \cite{T. Yamagata}, and then divided by the nucleon number of $^6$Li. }
\vspace{0.5cm}
\label{f0table}
\begin{tabular}{ccccccccc}
\hline
      Configuration   ~~  & $E_{b}$   & ~~   $E_\alpha$   &~~$E_{d(np)}$  &~~$E_{exp}$   \\
   (MeV/A)  & ~~ & ~~   & ~~    & ~~    \\ \hline
$\alpha + d$ &  ~~-3.49           &~~ -4.065            & ~~-2.34  &~~-5.087\\
 $\alpha + n + p$ &~~-3.26 	          & ~~-4.425     & ~~-0.93 &~~ -4.715 \\ \hline
\label{table}
 \end{tabular}
\end{table}

\subsection{GDR spectra of $^6$Li}

Before we show the GDR spectra, it is useful to see the time evolution of spatial separation between two centroids of neutrons and protons or of two clusters.
  For an example,  Fig.~\ref{fig_time} displays  the time evolution of the $\Delta_R(t)$ = $R_Z(t) - R_N(t)$ or $R_{\alpha}(t) - R_{d}(t)$
in coordinate space (Fig.~\ref{fig_time} (a)) and  their second derivatives of $\Delta_R(t)$ with respect to time (Fig.~\ref{fig_time} (b)) for  the $\alpha$+d configuration of $^6$Li.  In Fig.~\ref{fig_time}(a), we define the long axis as the $x$ axis,  then $x$-, $y$-
 and $z$-GDR represent three components of excited dipole moments along $x$, $y$ and $z$ directions of the oscillation of proton's core and neutron's core, and while  the blue long-dashed line (marked as ``x PDR" in insert) represents the oscillation between the core of $\alpha$ and the core of deuteron. It is cleanly seen that the frequencies of the oscillation between cores of protons and of neutrons are much higher than the one between the cores of $\alpha$ and of deuteron, which leads to a much higher energy of spectrum  of regular GDR than cluster resonance oscillation spectra as shown later.
 For its  second derivatives of $\Delta_R(t)$, the strength of cluster resonance between $\alpha$ and deuteron is also much weaker than all three components of regular GDR oscillations between neutron's core and proton's core.

\begin{figure}
\center
\includegraphics[scale=0.45]{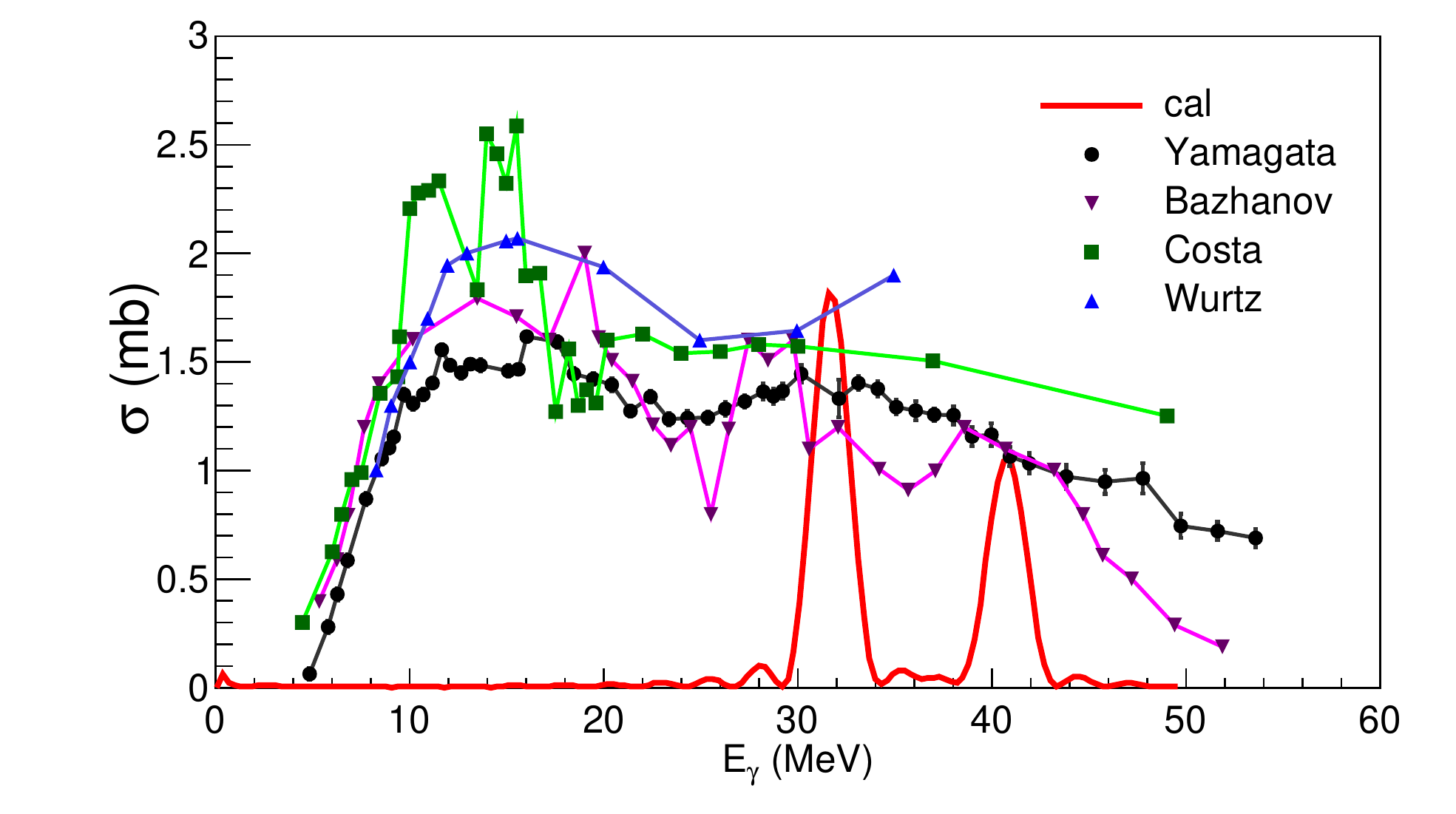}
\caption{
Comparison of GDR spectra of  our EQMD calculations assuming  $\alpha$ + deuteron (red line) structure for $^6$Li with the  experimental data. The peak around 31 MeV indicates of contribution from an $\alpha$ cluster inside $^6$Li  and while the one around 41 MeV could stem from the deuteron-like contribution inside $^6$Li.
The block dots with error bar, violet inverted triangles, green squares and blue triangles are various experimental  data from Yamagata {\it et~ al.}~\cite{T. Yamagata}, Bazhanov  {\it et~ al.}~\cite{Bathanov}, Costa  {\it et~ al.}~\cite{Costa}, and Wurtz~ {\it et~ al.} \cite{Wurtz}, respectively.
}
\label{fig_6Li-GDR}
\end{figure}

Fig.~\ref{fig_6Li-GDR} shows the calculation results of GDR spectra for $^{6}$Li,  in which the red
line  shows the result by the EQMD model with an assumption of $\alpha$ + deuteron structure for $^6$Li,  together with experimental data from several groups.  The first peak at about 31 MeV represents the GDR contribution from  the short axis, coming from the contribution of $\alpha$-cluster. This peak has been well discussed in previous work on $\alpha$-conjugated light nuclei, which has been taken as a fingerprint of $\alpha$-clustering structure inside a certain nucleus \cite{W.B.He,W.B.He2}.
  The second peak at 41 MeV can be attributed to the dominant  deuteron-like contribution due to neutron-proton pair. By comparing with the different experimental data obtained by Yamagata {\it et~al.} (block dot)~\cite{T. Yamagata},
Bazhanov {\it et al.} (violet) ~\cite{Bathanov}, Costa  {\it et al.} (green) ~\cite{Costa} as well as Wurtz {\it et al.}  (blue) ~\cite{Wurtz}, respectively, it seems that to the some extents the above two peaks from our calculations could explain some of  those data, especially of Yamagata {\it et~al.} and Bazhanov {\it et al.},  even though the data show much broad  area.  The data of Yamagata {\it et~al.}  show a very broad peak around 30 MeV and the data of Bazhanov {\it et al.} show a peak around 27 MeV, they could  stem from  a dipole oscillation of an $\alpha$ cluster inside $^6$Li because of the peak values are close to 31 MeV as predicted by He and Ma {\it et al.} \cite{W.B.He}.  On the other hand,   the data of Bazhanov {\it et al.} show another peak  around 40 MeV, which could be attributed to the  deuteron-like contribution inside $^6$Li.

\begin{figure}
\center
\includegraphics[scale=0.45]{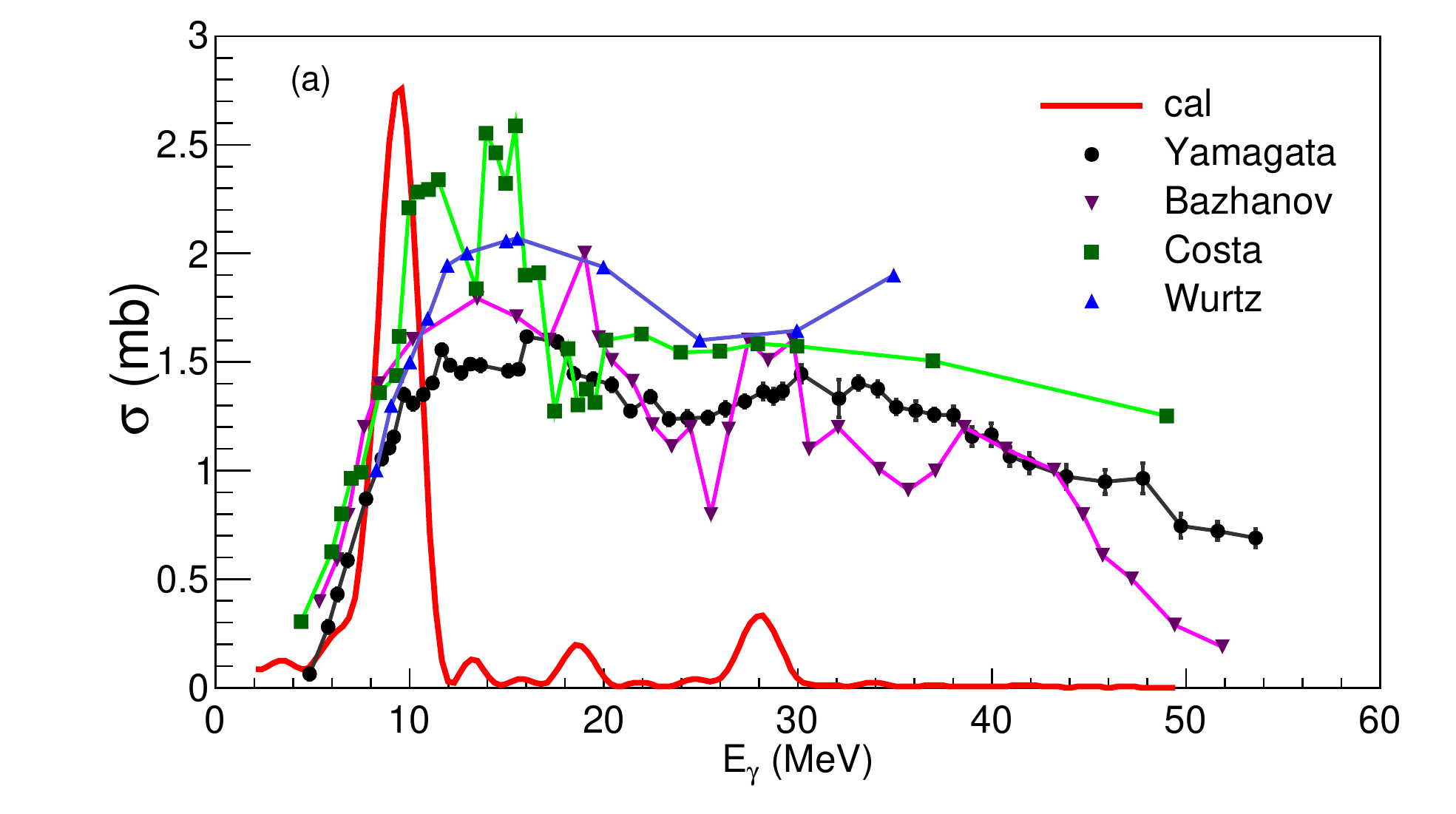}
\includegraphics[scale=0.45]{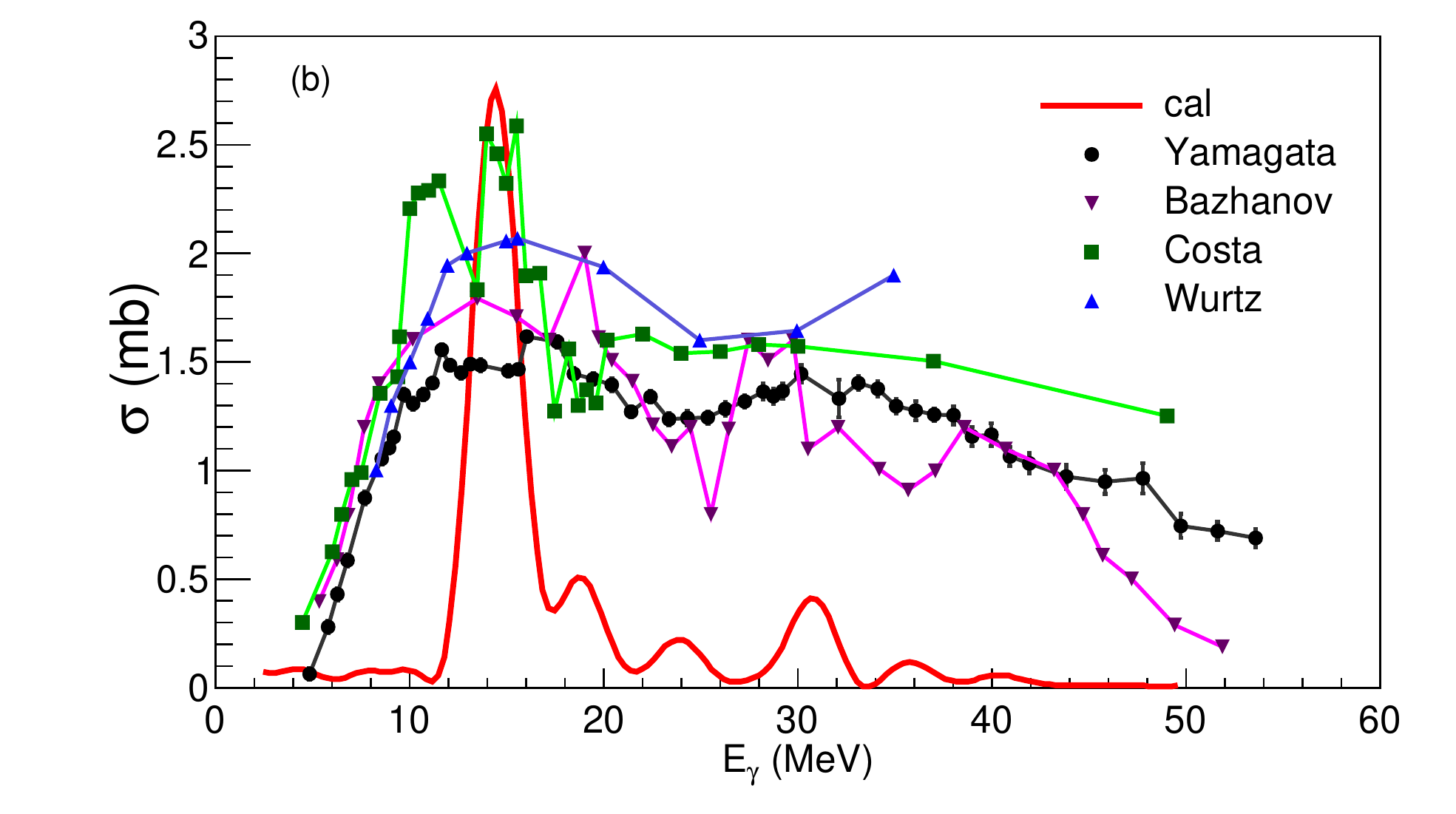}
\includegraphics[scale=0.45]{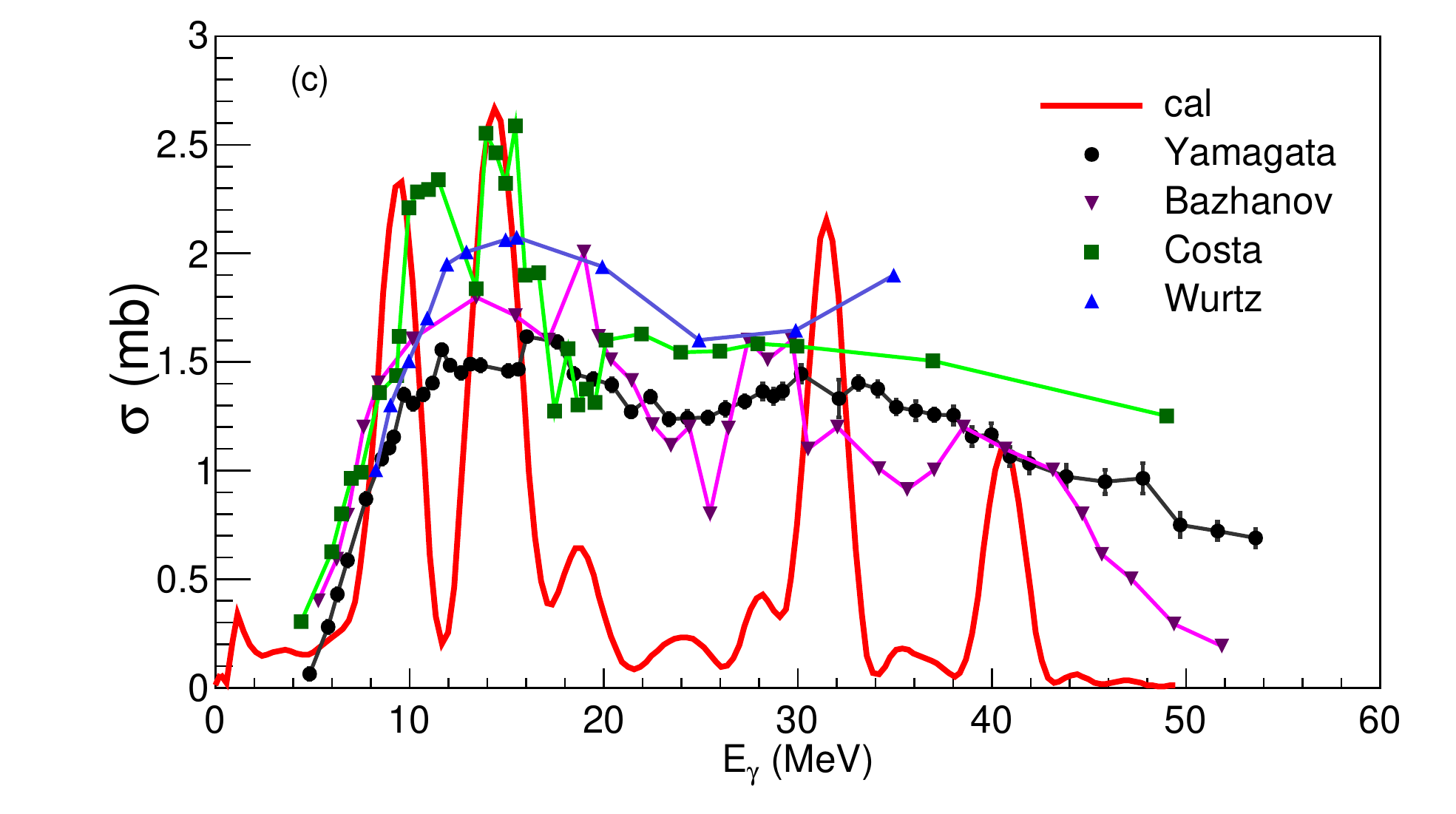}
\caption{Same as Fig.~\ref{fig_6Li-GDR} but for the
resonance oscillation calculations between $\alpha$ and $d$ (a) as well as  between $\alpha$  and  ($n$ + $p$) (b) for  $^6$Li.
Symbols are the data sets as Fig.~\ref{fig_6Li-GDR},  and  red lines are our EQMD calculations.
In panel (c), we perform a mixture  for four components, namely $\alpha$ + $d$ (i.e. panel (a)), $\alpha$ + $n$ + $p$ (i.e. panel (b)), $\alpha$-cluster peak around 31 MeV as well as  deuteron-like peak around 41 MeV (i.e. two separate peaks in Fig.~3) according to a  specific proportion of  1. : 1.17 : 0.9 : 1. }
\label{fig_6Li-GDR2}
\end{figure}

\begin{figure}
\center
\includegraphics[scale=0.45]{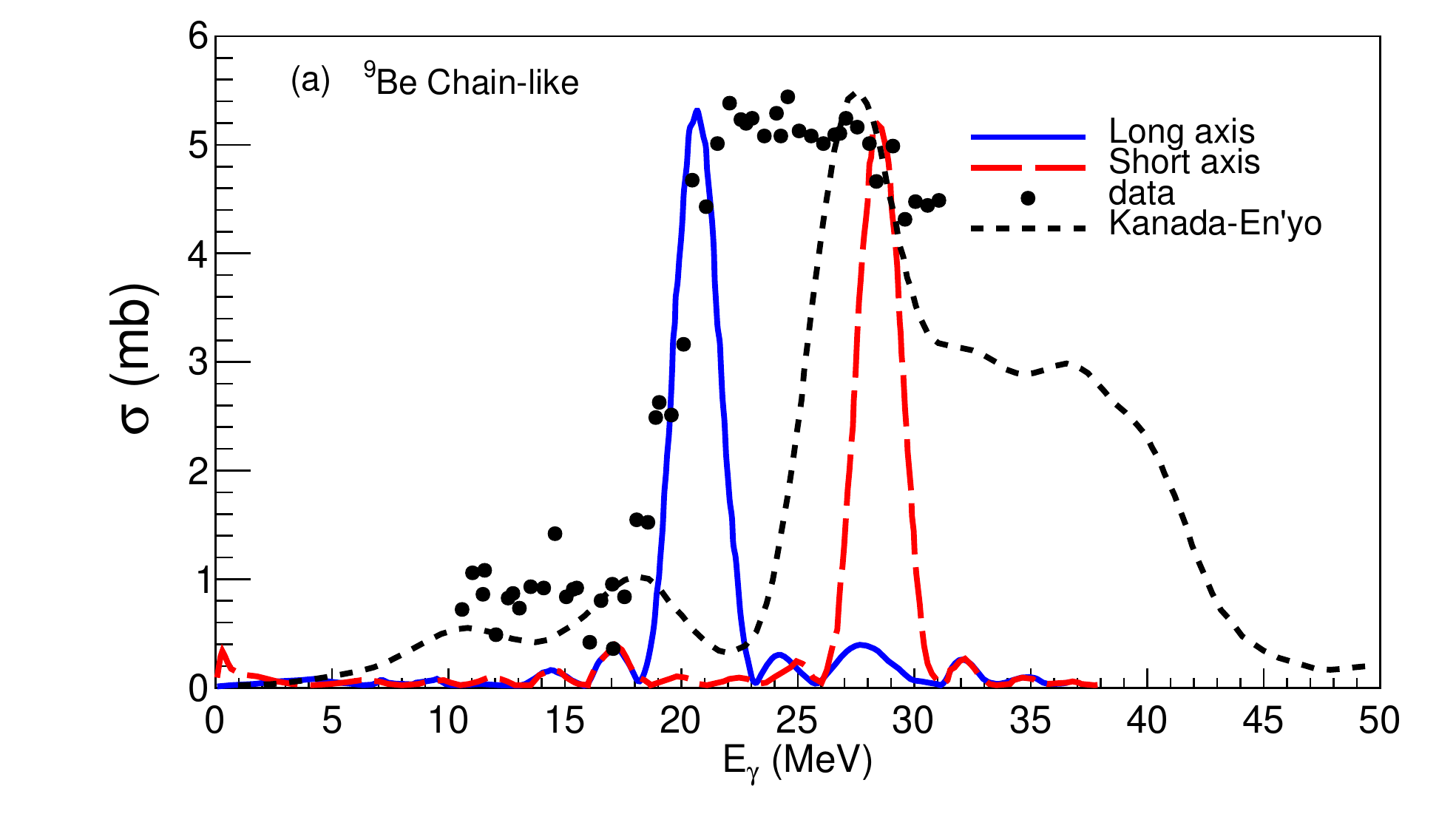}
\includegraphics[scale=0.45]{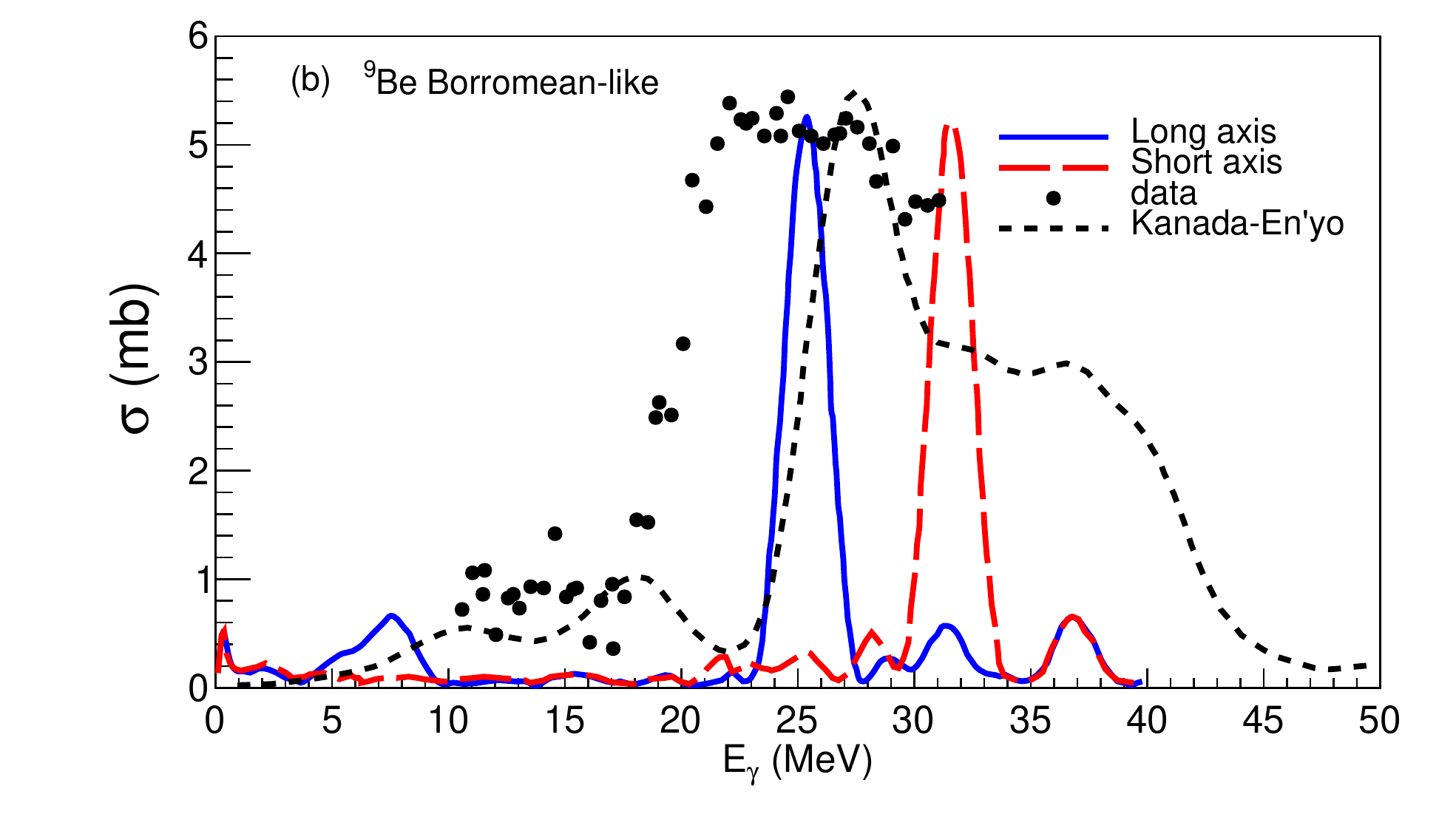}
\includegraphics[scale=0.45]{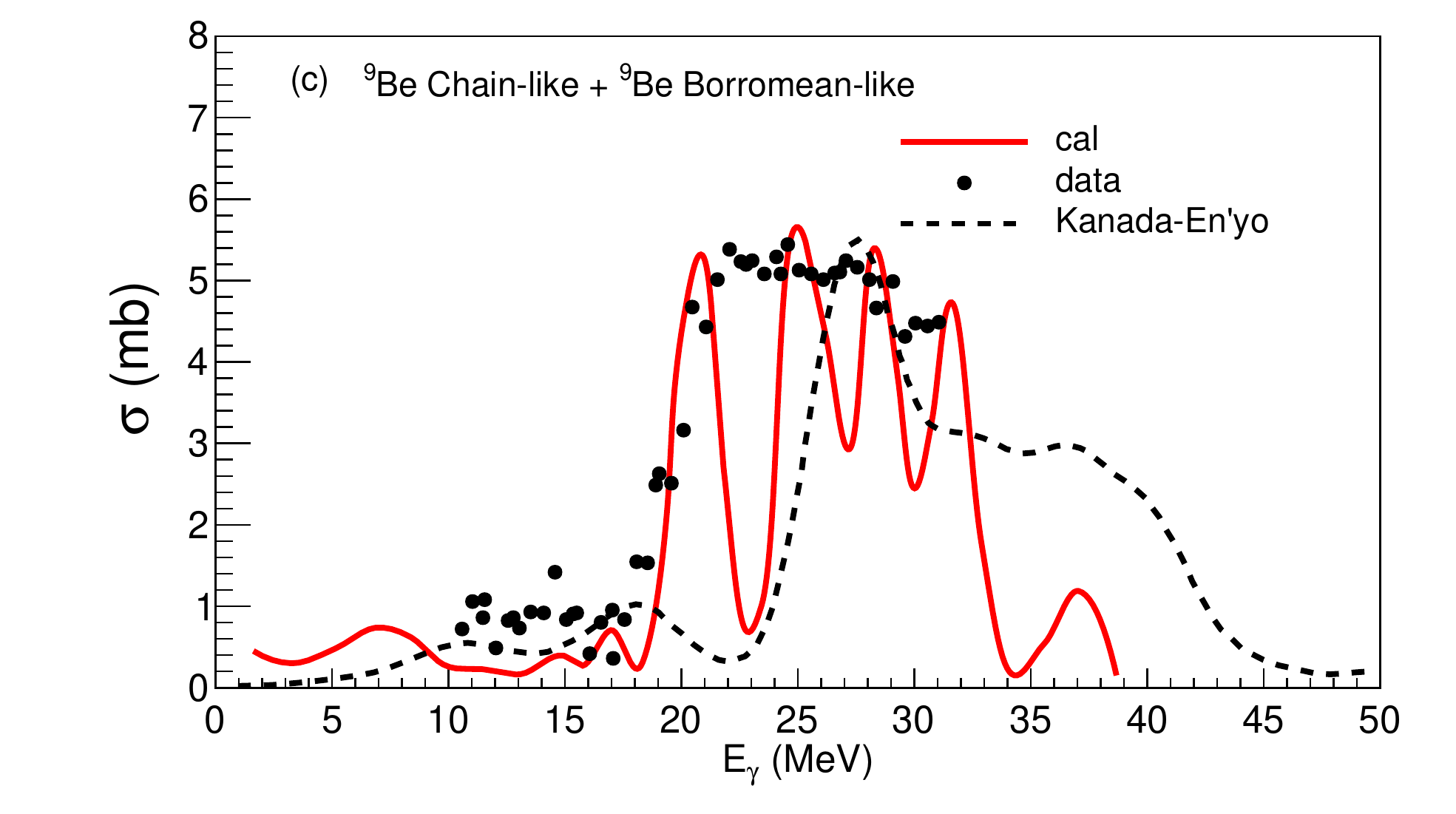}
\caption{
Comparison of GDR spectra of  our EQMD calculation assuming  the Chain-like  configuration (a)  as well as  the Borromean-like configurations (b) of three-body clustering structure of $\alpha$ + $n$ + $\alpha$ for $^9$Be with the  experimental data \cite{1975} as well as the sAMD calculation  \cite{Yoshiko}.
The black dots represent the experimental data via photonuclear reaction \cite{1975},  blue and red lines represent the present calculations,  and the dark  short dash line is the  sAMD calculation  result  from Ref.~\cite{Yoshiko}. 
In panel (c), we perform a mixture of the first and second peaks of Chain-like structure as well as those of Borromean-like structure according to a specific proportion of 1. : 0.9 : 1.1 : 0.75.}
\label{fig_Be9}
\end{figure}

For the first peak of  data from Yamagata {\it et al.}~\cite{T. Yamagata}, we assume that it is due to the collective cluster dipole resonance between the two clusters, namely $\alpha$ and deuteron. In order to check this assumption, we calculate the vibration spectrum between $\alpha$ and deuteron, i.e. through
  giant dipole moment between the centroids of the $\alpha$ and deuteron or the pair of neutron and proton. The corresponding  spectrum is  shown in Fig.~\ref{fig_6Li-GDR2}.
  Note that we scale the calculated strength of  dipole resonance   to a height similar to the experimental value  in order to compare with the experimental results intuitively. 
 The peak at about 10 MeV with the  red solid line in Fig.~\ref{fig_6Li-GDR2}(a)   is our result for cluster dipole resonance with  the structure of $\alpha$ + $d$, while the peak at around 14 MeV  with red line in Fig.~\ref{fig_6Li-GDR2}(b) represents  our result for the dipole resonance  spectra with the configuration of   $\alpha$ + $p$ + $n$.    
 We can say that the above double peaks around 10 and 14 MeV give an average main peak  at  about 12 MeV, which is close  the first peak of  data from Yamagata {\it et al.}, or we can say that the first broad peak of the data  could be attributed to mixture of resonance oscillations among clusters, i.e. $\alpha$ + $d$ and $\alpha$ + $p$ + $n$. 
On the other hand, for the data of Costa {\it et al.} (green squares), there are double peaks at lower photon energy, i.e. one is around 10 MeV and another is around 14 MeV,  it seems that our cluster resonances of $\alpha$ + $d$ and  $\alpha$ + $p$ + $n$ are consistent with the above double peaks, and  the $\alpha$ + $p$ + $n$ configuration might be a little more dominant. Nevertheless, no much significant difference in photon energies for this kind of  dipole resonances   is found between  $\alpha$ and $d$  or ($n$ + $p$), which could be seen as a kind of pygmy dipole resonance (PDR). 
   
   In general there is a coupling between different configurations, i.e. different configurations could be mixed for an overall contribution to GDR spectra.  To address  this question, 
 we perform a mixture of the $\alpha$ + $d$ (i.e. Fig.~\ref{fig_6Li-GDR2}(a)), $\alpha$ + $n$ + $p$ (i.e. Fig.~\ref{fig_6Li-GDR2}(b)), $\alpha$-cluster peak around 31 MeV and deuteron-like peak around 41 MeV (Fig.~\ref{fig_6Li-GDR}) according to a  specific proportion, eg. 1. : 1.17 : 0.9 : 1., respectively. The combined result is plotted in Fig.~\ref{fig_6Li-GDR2}(c). Even though the subtle structures, especially for widths are not fully described, our overall picture captures the characteristics of whole spectra of the experimental data. It needs to be pointed that the widths of each peak in our simulations are not well given due to the limit of our model in which  decay mechanisms of excited fragments are absent, and also the  specific proportion is model dependent.

\subsection{GDR spectra of $^9$Be}

Fig.~\ref{fig_Be9} shows the  dipole resonance   spectra calculation for the two types of three-body clustering structures of $^{9}$Be, one is the Chain-like structure (Fig.~\ref{fig_Be9} (a)) and another is the Borromean-like  configuration of $\alpha$ + $n$ + $\alpha$ (Fig.~\ref{fig_Be9} (b)) together with the earlier data \cite{1975} as well as sAMD calculation \cite{Yoshiko}. The Chain-like configuration shows double peaks, i.e. the one locates  at about 30 MeV (red dashed line), which is   due to the GDR of $\alpha$ cluster \cite{W.B.He,W.B.He2}, and the another lower energy peak at  around 20 MeV (blue solid line) is the contribution from component of the long axis. Similarly, for the Borromean-like  configuration as presented in Fig.~\ref{fig_Be9} (b), double-peak structure also  emerges, where
the high (red dashed line) and low (blue solid line) energy peaks indicate of the component of the $\alpha$ cluster and a component of the long axis, respectively, as the Chain-like case. For theoretical comparison, we also plot the calculation of Ref.~\cite{Yoshiko} which was renormalized to the height of the data for the main peak as shown in a dark short dash line, in which the first stronger peak comes from the main contribution by the longitudinal mode, and the second  higher energy broad peak can be attributed to the transverse mode for the configuration of $\alpha$ + $n$ + $\alpha$. It seems that, for the lower energy peak of Ref.~\cite{Yoshiko}, it is in agreement with the Borromean-like structure to some extents, but for its higher energy peak it is higher than our calculations for all the two configurations.
By comparing our calculations with the experimental data, it seems that the mixture of the Chain-like and the Borromean-like  three-body configurations of $\alpha$ + $n$ + $\alpha$ in our model calculation could describe the data  of $^{9}$Be. 
To this end, a mixture of different peaks is performed. We put the strengths of the first and second peaks of Chain-like structure as well as those of Borromean-like structure in terms of  1 : 0.9 : 1.1 : 0.75 to combine into a new figure,  i.e.  Fig.~\ref{fig_Be9}(c). Cleanly, the general structure of mixture peaks could more or less describe the experimental data even though the widths are not well described as we mentioned before.

\section{Summary}
Using different initial configurations through the cooling process in a framework of EQMD, we calculated the dipole resonance spectra for two non-$\alpha$-conjugated light nuclei, namely $^{6}$Li and $^{9}$Be,
by using the Goldhaber-Teller assumption.
For $^{6}$Li, both  initial configurations, i.e. two-body $\alpha$ + $d$ configuration as well as three-body $\alpha$ + $n$ + $p$  configuration, display a GDR component with a peak at around 31 MeV because of the fingerprint signal of $\alpha$-clustering structure inside the nucleus.
On the other hand, by comparing with experimental data,  a dipole resonance component with  a lower energy peak at about 12 MeV  might be assigned to the pygmy resonance between $\alpha$-cluster  and deuteron or ($n + p$) according to our calculation.
For  $^{9}$Be, our calculated GDR spectra indicate that there is a configuration mixture  for the Chain-like  and the Borromean-like  structure
of three-body structure of $\alpha$ + $n$ + $\alpha$. 
Furthermore, different GDR and PDR components are tentatively mixed for  trying to describe the whole spectra of $^6$Li and $^9$Be, even though quantitative fits are not achieved due to the limit of the present model,  the overall    characteristics of whole spectra of the experimental data are captured by this work.  It sheds light on the updated structure information of  $^{6}$Li and $^{9}$Be from the GDR spectra in terms of $\alpha$-clustering aspect.

\vspace{0.5cm} 
This work is partially supported by the National Natural
Science Foundation of China under Contracts Nos. 11905284, 11890710 and 11890714, 
the Strategic Priority Research Program of the CAS under Grants No. XDB34000000,  and the 
Guangdong Major Project of Basic and Applied Basic Research No. 2020B0301030008.

\end{CJK*}

\end{document}